\documentclass[twocolumn,showpacs,showkeys,preprintnumbers,amsmath,amssymb]{revtex4-1}

\newtheorem{defi}{Definition}

\usepackage[utf8]{inputenc}
\usepackage{graphicx}
\begin{document}
\title{The Vernon transform and its use in quantum thermodynamics}
\date{\today}

\author{Erik Aurell}
\email{eaurell@kth.se}
\affiliation{KTH -- Royal Institute of Technology, AlbaNova University Center, SE-106 91 Stockholm, Sweden}%

\author{Jan Tuziemski}%
\email{jan.tuziemski@fysik.su.se}
\altaffiliation[On leave from ]{Department of Applied Physics and Mathematics, Gdansk University of Technology}
\affiliation{Department of Physics, Stockholm University, AlbaNova University Center, Stockholm SE-106 91 Sweden}
\affiliation{Nordita, Royal Institute of Technology and Stockholm University,Roslagstullsbacken 23, SE-106 91 Stockholm, Sweden}

\begin{abstract}
The thermodynamics of a quantum system interacting with an environment 
that can be assimilated to a harmonic oscillator bath has been extensively
investigated theoretically. In recent experiments, the system under study
however does not interact directly with the bath, but though a cavity
or a transmission line. The influence on the system from the bath is therefore
seen through an intermediate system, which modifies the characteristics
of this influence.
Here we first show that this problem is elegantly solved by a transform,
which we call the Vernon transform,
mapping influence action kernels on influence action kernels.
We also show that the Vernon transform takes a particularly
simple form in the Fourier domain, though it then must be interpreted with some care.
Second, leveraging results in
quantum thermodynamics we
show how the Vernon transform can also be
used to compute the generating function
of energy changes in the environment.
We work out the example of a system interacting
with two baths of the Caldeira-Leggett type,
each of them seen through a cavity.
\end{abstract}

\maketitle

\section{Introduction}
The study of of heat released to or absorbed from a bath (or baths)
by a driven quantum system took off with the path-breaking
contribution by Alicki now 40 years ago~\cite{Alicki79}.
For recent reviews of the context and later developments, see~\cite{KosloffLevy2014,VinjanampathyAnders2016}.
In Alicki's approach the dynamics of a system is modeled as
a quantum Markov process \cite{Weiss-book,Breuer2002,Alicki-Lendi-book}, and energy exchange between the
system and the bath (or baths) are expressed in the Lindblad operators.
The quantum heat, defined as expected energy change in the bath, is then
the reverse of the expected dissipative energy change of the system,
which is determined by the Lindblad operators acting on the system quantum state.
\\\\
The setting has been extended in several directions, both recently
and less recently.
First, higher moments or the entire distribution of bath energy changes
may be of interest. They are not necessarily the same as higher moments
and distribution of system energy changes, but at least in a formal
sense these quantities remain quantum functionals
of the system history \cite{Esposito2009,AurellDonvilMallick2020,AurellKawaiGoyal2020}.
When modelling the dynamics of a qubit in a supercomputing circuit
as very low temperature \cite{Devoret1995,Wendin2017},
the quantum Markov process assumption is questionable.
Indeed, a wealth of phenomena have been worked out for
the problem of a qubit interacting with a bath
when the drive of the qubit changes on a time scale comparable to
or faster than the bath \cite{GrifoniHanggi1998}.
Other types of explicit results have been obtained in the
spin-boson problem~\cite{Leggett87},
in the "non-interacting blip approximation" (NIBA).
The relaxation time scale of the bath is then assumed shorter than the times
between system jumps, but not zero, and far longer than the
time over which a jump a takes place
\cite{SegalNitzan2005,Segal2008,Segal2006,AurellMontana,AurellDonvilMallick2020}.
In a related direction, when the interaction between the
system and the bath is strong, meaning that 
in a process the typical energy 
stored in system-bath interactions is comparable to the variations
in system energy, the concept of heat is delicate even classically,
see 
\textit{e.g.}~\cite{Seifert2016,TalknerHanggi2016,Jarzynski2017,MillerAnders2017,Aurell2017}.
In the quantum domain these questions have been actively 
investigated by many groups with different techniques
\cite{EspositoOchoaGalperin,Kato2015,Carrega2015,AurellEichhorn2015,Carrega2016,Kato2016,Newman2017,Motz_2018,Aurell2018b,Douetal2018,PernauLlobet2018,FunoQuan2018,Kwonetal2018}.
\\\\
Our goal here is a different one, and motivated by 
experimental set-ups when investigating heat flow 
through superconducting qubits~\cite{Ronzani2018}.
In such devices the bath (or baths) with which the qubit
eventually exchanges energy are (small) normal-metal components held at
fixed temperature. The influence of such baths, which are 
physically comprised of a relatively small number of
conduction-band electrons (fermions) excited above the ground state
may be assimilated to a bath of harmonic oscillators that
would classically act as friction/resistance~\cite{Donvil2018}.
However, in the experimental set up, this (these) bath(s) 
do not interact directly with the qubit, but through a 
single-mode transmission line.
The relevant physical model for most experiments
on quantum thermodynamics is hence that of
(system)-(cavity)-(bath).
\begin{figure}
\centering
\includegraphics[scale=0.15]{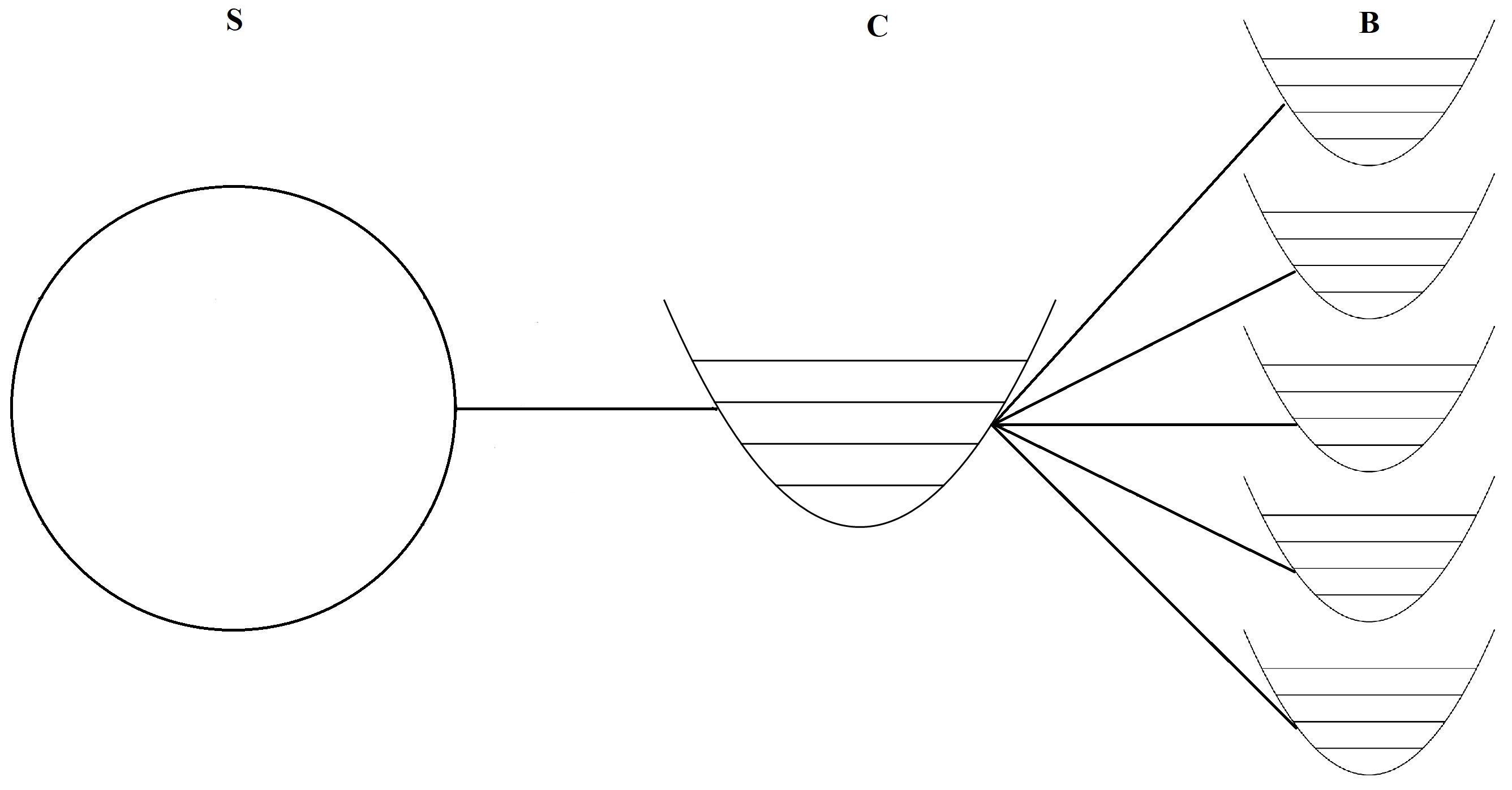}
		\caption{\label{fig:SOB} 
A schematic illustration of the set-up considered in this paper.
A system of interest (S) interacts with a cavity (C) which interacts 
with an environment (B).
It is assumed that (B) is described by the kernels of the
real and imaginary parts of the Feynman-Vernon action,
$k_I^{b\to c}$ and $k_R^{b\to c}$. This means that whatever its actual physical
constituents, as to its influence on C, the environment B behaves 
as a harmonic oscillator bath.
C is described as one oscillator degree of freedom (a mode in an actual cavity,
or a transmission line) with mode frequency $\omega$, which
interacts linearly with both (S) and (B).
The influence on (S) from its total environment,
(C) and (B) considered together, is then described Feynman-Vernon
action kernels
$k_I^{c\to s}$ and $k_R^{c\to s}$.
The \textit{Vernon transform} maps $(k_I^{b\to c},k_R^{b\to c})$
on $(k_I^{c\to s},k_R^{c\to s})$.
It is described in text how this is used to express quantum heat
(energy changes in the bath) as functionals of system history.
 }
\end{figure}
Even if it is valuable to know how a given bath influences a given system
through a direct linear coupling
this is hence not always directly applicable. 
The system only experiences the bath as seen through (filtered by)
the oscillator, and this changes the effects on the system
dramatically. While several theoretical investigations have been
performed, \textit{e.g.} recently in
\cite{Meng2021} and \cite{PekolaKarimi2020},
we believe the perspective taken here is sufficiently
different to motivate another investigation on the topic.
\\\\
In particular, outside the 
specialized literature it does not seem to be generally appreciated that
\textit{the bath seen through the oscillator constitutes
a systematic transformation of the bath}.
It is further not generally known that this transformation
was written down by F.~Vernon 
in what may be the very first publication on open quantum systems,
his 1958 Caltech PhD thesis~\cite{VernonPhD}.
Vernon's supervisor was Feynman, and much of the material in Vernon's
thesis can also be found in their famous
later joint paper~\cite{FeynmanVernon1963}.
However, the transformation which will be the
main tool in the following
is not found in~\cite{FeynmanVernon1963}.
In Vernon's honor we thus refer to
it as the \textit{Vernon transformation}.
\\\\
Combining recent development in quantum thermodynamics and
the Vernon transform, 
we will further show that we can compute the
generating function of energy change in the
total environment, \textit{i.e}
the bath and the cavity considered together.
We will show that these formula are particularly transparent
for the first moment that gives quantum heat, and
further simplifies for \textit{quantum power}, the quantum heat per unit time.
They can be given in almost closed for when the bath
is of the Caldeira-Leggett model~\cite{CALDEIRA1983}.
Although technically more involved,
we can also compute moments 
of the energy changes in the bath only;
these details are presented in an appendix.
\\\\
The paper is organized as follows.
In Section~\ref{sec:Vernon} we introduce the model 
of a system, an oscillator and bath, the oscillator interacting
linearly with both the system and the bath,
and we derive the Vernon transform
for the real part of the Feynman-Vernon action kernel.
Details are given in
Appendices~\ref{sec:vernon-feynman-vernon}
and~\ref{sec:vernon-real}.
Thermodynamics and generating function of quantum heat
in a system interacting to more than one bath
are introduced in Section~\ref{eq:thermo}.
In Section~\ref{sec:Caldeira-Leggett} we apply this 
method to a system interacting with two cavities,
each of them interacting with its own bath as
in the Caldeira-Leggett
model. In Section~\ref{sec:discussion}
we sum up and discuss our results.
In Appendix~\ref{app:vernon-transform-heat-bath-alone} we discuss 
how to estimate
the energy deposited in the bath only (not in the cavity).
In that case we can only compute moments of the energy 
change and not the complete generating function,
and in practice only the first moment.
A bath initially 
in equilibrium with constant, time-independent interactions
between the bath and the system
has issues
previously addressed by
Caldeira and co-workers~\cite{daCosta2000}.
and H\"anggi and co-workers~\cite{Ingold2009}.
For completeness we summarize these issues in Appendix~\ref{sec:problems}.

\section{The Vernon transform}
\label{sec:Vernon}
We are interested in Hamiltonian operators of the form
\begin{eqnarray}
\label{eq:total-Hamiltonian}
\hat{H}(t)= \hat{H}_S+\hat{H}_C+\hat{H}_B + \hat{H}_{SB}(t) + \hat{H}_{BC}(t),
\end{eqnarray}
where $H_S$ is a system Hamiltonian, possibly time-dependent, 
the bath 
is a collection of harmonic oscillators $H_B = \sum_k \omega_k b_k^\dagger b_k$
and the cavity consists of one 
harmonic oscillator $H_C = \omega_C a^\dagger_C a_C$.
Initially the bath is in a state of
thermal equilibrium with respect to $H_B$.
The cavity interacts linearly with both the bath and the system.
In the following discussion it will at some points be
convenient to allow these interactions to be time-dependent,
so that we have 
\begin{eqnarray}
\hat{H}_{SB}(t) &=& C_{SC}(t)\, \hat{Q}_S \otimes \hat{X}_C \\
\hat{H}_{BC}(t) &=& C_{CB}(t)\, \hat{X}_C \otimes \sum_k \hat{Y}_B
\end{eqnarray}
with possibly time-dependent coefficients $C_{SC}(t)$
and $C_{CB}(t)$, and where 
$\hat{Y}_B = \sum_k \frac{1}{2} \left( \hat{b}_k +\hat{b}_k^\dagger \right)$.
It is well known that in this situation
the influence of the bath on the cavity is completely described
by the \textit{Feynman-Vernon action}~\cite{FeynmanVernon1963,Weiss-book,Breuer2002}.
In path integral language this is a quadratic functional
of forward and backward \textit{cavity paths}, representing unitary evolution 
operators $U$ and $U^{\dagger}$ acting on the cavity after tracing out the bath.
It is also well known that the influence on the system of the cavity and the bath
together is described by another 
Feynman-Vernon action, a quadratic functional
of forward and backward \textit{system paths}.
The \textit{Vernon transform}~\cite{VernonPhD} (Appendix 5)
expresses the kernels of the second Feynman-Vernon action in terms of the 
kernels of the first. In this way one sees how the system experiences
the bath when its influence is transmitted through the cavity.
\\\\
We start by stating an intermediate result
in Vernon's derivation of the Feynman-Vernon action itself.
For convenience we repeat that derivation in
Appendix~\ref{sec:vernon-feynman-vernon}.
The central observation is that 
a double path integral for a harmonic oscillator
can be re-written in terms of the
of the sums and differences of forward and backward paths,
and thus gives rise to an auxiliary function
satisfying
$$\Delta y^*_f=\Delta\dot{y}^*_f=0\quad \Delta \ddot{Y}^* = - \omega_k^2 \Delta Y^* 
+ C_{CB}(t)\Delta X=0 $$
This is the equation of motion of
a harmonic oscillator 
with an external drive starting from rest at 
the final time $t_f$, and 
evolving backwards in time to 
the initial time $t_i$.
The value of $\Delta Y^*$ at time $t$ hence depends
on the values of $\Delta X$ at times $s$ larger than $t$,
but not on the 
the values of $\Delta X$ at times $s$ less than or equal 
to $t$.
The Feynman-Vernon influence functional 
from one bath oscillator on the cavity is then
\begin{eqnarray}
\label{eq:Vernon-Feynman-Vernon}
\mathcal{F}^{k\to C} &=&
e^{\frac{i}{\hbar}\int_{t_i}^{t_f} 
\frac{C_{CB}(t)}{2}\bar{X}\Delta Y^*\, dt} \nonumber \\
&& \cdot e^{-\frac{1}{4\hbar}
\hbox{coth}\left(\frac{\omega_k\beta\hbar}{2}\right)
\left(\frac{1}{\omega_k} \left(\Delta\dot{y}^*\right)^2
+ \omega_k\left(\Delta{y}^*\right)^2\right)}
\end{eqnarray}
The dependence of 
$\Delta Y(t)$ on $\Delta X(s)$
can be written
\begin{equation}
\label{eq:response}
\Delta Y^*(t) = \int_t^{t_f} R_k(t,s) C_{CB}(s) \Delta X(s) ds
\end{equation}
where $R_k=\frac{1}{\omega_k} \sin\omega_k (s-t) \Theta(s-t)$ is the response function of a harmonic oscillator.
Inserting this in
the imaginary term in the exponent
\eqref{eq:Vernon-Feynman-Vernon}
it becomes a double integral
\begin{equation}
\label{eq:S_i-k-to-C}
S_i^{k\to C} = \int_{t_i}^{t_f}\int_t^{t_f} 
\frac{C_{CB}(t)C_{CB}(s)}{2}\bar{X}(t) R_k(t,s) 
 \Delta X(s)\, dt\, ds
\end{equation}
which is the standard form of the real part of the
Feynman-Vernon action from one bath oscillator (see below).

Now we consider the system-cavity-bath situation
and assume that the cavity starts in equilibrium with
respect to $\hat{H}_C$. The influence of 
all the bath oscillators on the
cavity is expressed as the real and imaginary parts of the
bath-cavity Feynman-Vernon action
\begin{eqnarray}
\label{eq:S_i-B-to-C}
S^{B\to C}_i&=& \frac{i}{\hbar}
\iint^{t_f,t} k_i^{B\to C}(t,s)\Delta X(t)\bar{X}(s)
\, ds\, dt  \\
&& k_i^{B\to C}=\sum_k\frac{C_{CB}(t)C_{CB}(s)}{2\omega_k}\sin\omega_k(t-s) \nonumber \\
\label{eq:S_r-B-to-C}
S^{B\to C}_r&=& ´-\frac{1}{2\hbar}
\iint^{t_f} k_r^{B\to C}(t,s)\Delta X(t)\Delta X(s)
\, ds\, dt  \\
 k_r^{B\to C}&=&\sum_k\frac{C_{CB}(t)C_{CB}(s)}{2\omega_k}\cos\omega_k(t-s) \hbox{coth}\frac{\omega_k\hbar\beta}{2} \nonumber
\end{eqnarray}
The real action kernel ($S^{B\to C}_i$) has been derived above
and the imaginary action kernel ($S^{B\to C}_r$) is derived
in Appendix~\ref{sec:vernon-feynman-vernon}.
We are here interested in the real action, as only this one
depends on $\bar{X}$.

The problem of integrating the bath and the cavity
is solved by a new auxiliary function
which satisfies
\begin{eqnarray}
\Delta \ddot{X}^* + \omega_C^2 \Delta X^* 
&=& C_{SC}(t)\Delta Q + 2\int_t^{t_f} k_i^{B\to C}(s,t) \Delta X^*(s) \nonumber \\
\label{eq:Vernon-auxiliary-2}
\Delta x^*_f=\Delta\dot{x}^*_f&=&0
\end{eqnarray}
In above
$\Delta Q$ is the difference of the forward 
and backward paths of \textit{the system}, and 
$C_{SC}(t)$ is the \textit{system-cavity coupling}.
The equation is again that of 
a harmonic oscillator 
with an external drive starting from rest at 
the final time $t_f$, and 
evolving backwards in time to 
the initial time $t_i$.
However, there is now also a damping term
which contains the effects of the bath.
It is well known that for an Ohmic bath
(Caldeira-Leggett model), the
kernel $k_i^{B\to C}(s,t)$ is 
$-\eta \dot{\delta}(s-t)$ where $\eta$
is a classical friction coefficient;
the damping integral 
\eqref{eq:Vernon-auxiliary-2}
is then an ordinary friction term 
$-\eta \Delta\dot{X}^*$. 
The example is treated in
Section~\ref{sec:Caldeira-Leggett}. 

Whatever the influence from the bath, the
value of $\Delta X^*$ at time $t$ depends
on the values of $\Delta Q$ at times $s$ larger than $t$.
We can write that as
\begin{equation}
\label{eq:response-2}
\Delta X^*(t) = \int_t^{t_f} R_C(t,s) C_{SC}(s) \Delta Q(s) \, ds
\end{equation}
where $R_C(t,s)$ is the response function of the
damped harmonic oscillator describing the cavity.
The real Feynman-Vernon action of the cavity and the
bath on the system is hence expressed as as double
integral analogous to \eqref{eq:S_i-k-to-C}:
\begin{equation}
\label{eq:S_i-C-to-S}
S_i^{C\to S} = \int_{t_i}^{t_f}\int_t^{t_f} 
\frac{C_{SC}(t)C_{SC}(s)}{2}\bar{Q}(t) R_C(t,s) 
 \Delta Q(s)\, dt\, ds
\end{equation}
We are now in a position to state
the Vernon transform.
For clarity we do it first for the general
case and then (below) under simplifying
assumptions and in the Fourier domain.

\begin{defi} \label{def:Vernon-transform-1}
The Vernon transform ${\cal V}$ of a
Feynman-Vernon kernel 
$k_i^{B\to C}$ on another
Feynman-Vernon kernel 
$k_i^{C\to S}$ is given by
\begin{equation}
\label{eq:Vernon-transform-def}
k_i^{C\to S}(t,s)={\cal V}\left[k_i^{B\to C}\right]=\frac{C_{SC}(t)C_{SC}(s)}{2} R_C(t,s) 
\end{equation}
where the cavity response function is defined
by
\eqref{eq:response-2}
and the auxiliary function $\Delta X^*$
satisfies \eqref{eq:Vernon-auxiliary-2}.
\end{defi}

The Vernon transform is especially convenient on
the Fourier side, and when assuming that all 
interactions are time-independent and
the process goes on for all time.
There are physical issues with such a model
which we discuss in
Appendix~\ref{sec:problems}, but in this
paper we will mostly leave these aside.
The Feynman-Vernon kernel $k_i^{B\to C}(t,s)$
then only depends on the time difference 
$\tau=s-t$ and is represented by its Fourier 
transform
\begin{equation}
\hat{k}_i^{B\to C}(\nu)=\int_{-\infty}^{\infty}
e^{i\nu\tau} k_i^{B\to C}(\tau) \, d\tau
\end{equation}
The auxiliary function $\Delta X^*$
is hence Fourier domain
given by
\begin{eqnarray}
\Delta \hat{X}^*(\nu)
&=& \frac{C_{SC}\Delta \hat{Q}(\nu)}
{-\nu^2 + \omega_C^2 - 2\hat{k}_i^{B\to C}(-\nu)}
\label{eq:Vernon-auxiliary-3}
\end{eqnarray}
where the denominator is the response function
$\hat{R}_C$ in the Fourier domain.
From this we have

\begin{defi} \label{def:Vernon-transform-2}
The Vernon transform ${\cal V}^{\infty}$ 
on an infinite time interval of a
Feynman-Vernon kernel 
$k_i^{B\to C}$ which only depends on the 
time difference and where all interactions
are time-independent
is in the Fourier domain given by
\begin{equation}
\label{eq:Vernon-transform-Fourier-side}
\hat{k}_i^{C\to S}(\nu)
=\frac{1}{2}\frac{C_{SC}^2}
{-\nu^2 + \omega_C^2 - 2\hat{k}_i^{B\to C}(-\nu)}
\end{equation}
\end{defi}

The Vernon transform is non-linear. For this reason we separate it from the 
analogous mapping of the real kernels which is linear
in the real kernels, and which we discuss in Appendix~\ref{sec:vernon-real}.
For completeness we state it here on the Fourier side as
\begin{defi} \label{def:Vernon-transform-real}
The real Vernon transform ${\cal W}$ 
is under assumptions of
Definition~\ref{def:Vernon-transform-2}
and excepting boundary terms
a linear mapping of the
Feynman-Vernon kernel
$\hat{k}_r^{B\to C}(\nu)$ on the
Feynman-Vernon kernel
$\hat{k}_r^{C\to S}(\nu)$ given by
\begin{equation}
\label{eq:Vernon-transform-Fourier-side-real}
\hat{k}_r^{C\to S}(\nu)
= \frac{1}{2} C_{SC}^2 \, \hat{k}_r^{B\to C}(-\nu)\,  \hat{R}_C(\nu) \hat{R}_C(-\nu)
\end{equation}
where $\hat{R}(\nu)$ is the Fourier transform of the response
function. The real Vernon transform ${\cal W}$ 
hence depends quadratically on $R_C$.
\end{defi}

\section{The Vernon transform in quantum thermodynamics}
\label{eq:thermo}
In the previous Section 
we showed how the system state is influenced by a bath when the interaction between them is mediated via the cavity. In this Section we are interested in thermodynamics of such setups. 
The central quantity of
interest is heat flow through a system
between reservoirs.
The set-up in this section will 
hence be that of a system 
coupled to two baths with
one cavity in between on 
each side.

\begin{figure}
\centering
\includegraphics[scale=0.1]{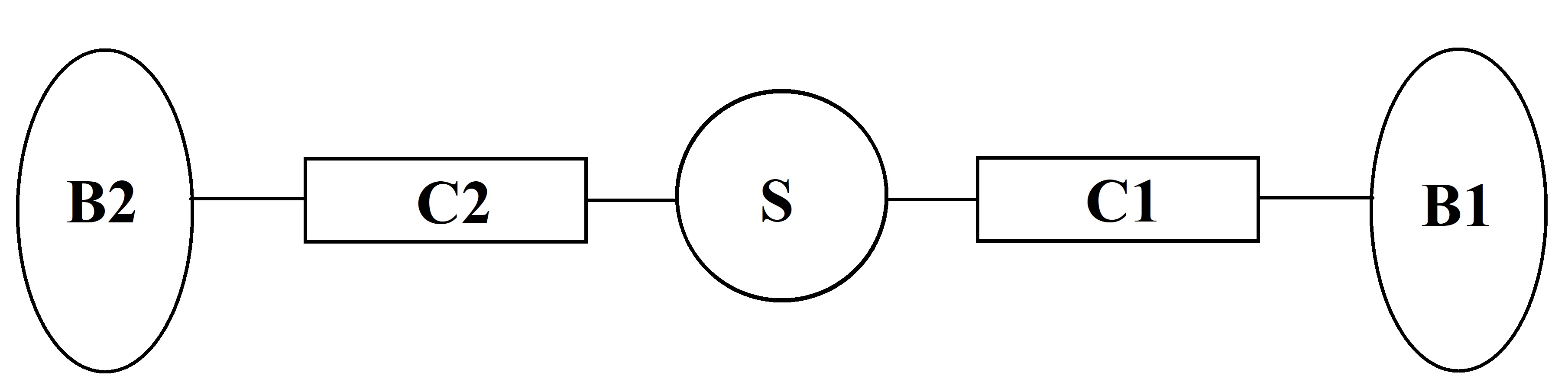}
		\caption{\label{fig:schematic-Ronzani} 
A schematic illustration 
of heat flow between reservoirs.
A system of interest (S) interacts with 
two cavities (C1 and C2) which interact 
with two environments (B1 and B2).
It is assumed that (B1 and B2) are described by the kernels of the
real and imaginary parts of the Feynman-Vernon action,
$k_I^{B1\to C1}$, $k_R^{B1\to C1}$, 
$k_I^{B2\to C2}$ and $k_R^{B2\to C2}$. 
This means that whatever their actual physical
constituents, as to their influence on 
the cavities, the environments behave 
as harmonic oscillator baths.
C1 and C2 are described as oscillator degrees of freedom with mode frequencies 
$\omega_{C1}$ and $\omega_{C2}$ which
interact linearly with the system.
The influence on (S) from its total environment
consists of two parts, one 
from (C1) and (B1), and one from 
(C2) and (B2), each described by a 
pair of Feynman-Vernon
action kernels
$(k_I^{C1\to S},k_R^{C1\to S})$
and 
$(k_I^{C2\to S},k_R^{C2\to S})$.
$k_I^{c\to s}$ and $k_R^{c\to s}$.
As discussed in text we make the simplifying
assumption that heat 
given to side 1(2) is the total energy 
change of C1 and B1 (C2 and B2).
The more involved case where heat given to
side 1(2) is the energy change in B1
only (B2 only) is treated in Appendix~\ref{app:heat-to-bath-alone}.
 }
\end{figure}
Conceptually we are faced with the question
what is heat. A natural definition would be to say
that heat on side 1(2) is the energy change in bath 
B1(B2), while energy change in cavity C1(C2) counts
as a kind of interaction energy which is not
necessarily lost to the system, but which can be
returned and do useful work.
The generating function of such a bath energy change
is analogous to the generating
function of a system interacting 
directly a bath which has been discussed 
multiple times {\it e.g.}
\cite{Esposito2009,Andrieux2009,Garrahan2010,Nicolin2011b,Nicolin2011,Aurell2018b,AurellKawaiGoyal2020,AurellDonvilMallick2020}, and other papers cited therein.
For the simpler one-sided case it reads

\begin{eqnarray}
G_{i f}(\nu)&=&\operatorname{Tr}_{CB}
\langle f| e^{i \nu H_{B}}U(t) e^{-i \nu H_{B}} \nonumber \\&& \times \left(\rho_{C} \otimes \rho_{\beta_{B}} \otimes|i\rangle\langle i| \right) U^{\dagger}(t) |f\rangle, 
\label{eq:Gif-nu}
\end{eqnarray}
where $|i\rangle$ ($|f\rangle$) denotes the initial (final) state of the system, $\rho_{C}$ and $\rho_{\beta_{B}}$ are initial states of the cavity and the bath (one each), respectively, and $U(t)$ is the evolution of the system, cavity, and the bath generated by the Hamiltonian 
\eqref{eq:total-Hamiltonian}.
It turns out that while moments of this quantity
can be computed, see
Appendix~\ref{app:heat-to-bath-alone},
the resulting expressions 
in the system variables, after integrating out the cavity,
contain several terms, and are not 
very transparent. The full generating function
\eqref{eq:Gif-nu}
can also not be simply expressed 
in the Vernon transform, which is the theme of this paper.

In the following we have instead taken the route of defining heat 
given to side 1(2) to be the total energy change of the
cavity and the bath on side 1(2). In the long time
limit, and barring the case where energy is built
up in the cavity this should to leading order give
the same behaviour as counting heat in the bath only.
The generating functions we will consider are
thus
\begin{eqnarray}
G^T_{i f}(\nu_1,\nu_2)&=&\operatorname{Tr}_{CB}
\langle f| e^{i \nu_1 (H_{B1}+H_{C1})
+i \nu_2 (H_{B2}+H_{C2})} \nonumber \\
&& U(t) e^{-\nu_1 (H_{B1}+H_{C1})
- \nu_2 (H_{B2}+H_{C2})} \nonumber \\&& \times \left(\rho_{C} \otimes \rho_{\beta_{B}} \otimes|i\rangle\langle i| \right) U^{\dagger}(t) |f\rangle, 
\label{eq:Gif-nu-total}
\end{eqnarray}
where $\operatorname{Tr}_{CB}$ stands for the trace 
of both baths and cavities,
$\rho_{C} \otimes \rho_{\beta_{B}}$ stand for the
initial product state of both baths and cavities,
and where we have assumed that the cavity-bath interaction
vanishes at the beginning and the end of the process.
From general results obtained in 
\cite{AurellKawaiGoyal2020} it follows
that  the  generating  function
\eqref{eq:Gif-nu-total}
is given  by a modified Feynman-Vernon action
\begin{eqnarray}
G_{if}(\kappa_1,\kappa_2)&=&\int_{if}  {\cal D} Q {\cal D} Q' 
   e^{\frac{i}{\hbar} \int_{t_i}^{t_f} dt \left( S_S[Q]-S_S[Q']
   \right)} \rho_S\left(Q,Q' \right) \nonumber \\ 
&&    
   \mathcal{F}^{C1 \to S}_{\kappa_1}[Q, Q'] \mathcal{F}^{C2 \to S}_{\kappa_2}[Q, Q']    
   \label{eq:G-nu-def}
\end{eqnarray}
where the only change in 
is the substitution of
the Feynman-Vernon functional
$\mathcal{F}^{C\to S }[Q, Q']$
(either side)
with
\begin{eqnarray}
    \begin{aligned}
\mathcal{F}^{C\to S}_{\kappa}[Q, Q'] &=e^{\frac{i}{\hbar}S_{i, \kappa}^{C1\to S}[Q, Q']-\frac{1}{\hbar} S_{r, \kappa}^{C1\to S}[Q, Q']}
\end{aligned}
\end{eqnarray}
For simplicity we will
in the following drop $\kappa_2$ and only consider the
generating function of energy changes on side 1 only.
The imaginary and real part of the modified Feynman-Vernon 
cavity-to-system action depend on $\kappa$. 
For the case when the bath-cavity and cavity-system
coupling
coefficients constant they read
\begin{eqnarray}
S_{i, \kappa}^{C1\to S}[Q, Q'] &=& \int_{t_{i}}^{t_{f}} \mathrm{~d} t \int_{t_{i}}^{t} \mathrm{~d} s\Big[ k_{i}^{C1\to S}(t-s) \big(Q(t) Q(s)- \nonumber \\
&& Q'(t) Q'(s)\big)  +   k_{i}^{C1\to S}\left(t-s+\kappa\right) Q(t) Q'(s)
\nonumber \\
&& \quad-\,k_{i}^{C1\to S}\left(t-s-\kappa\right)Q(s) Q'(t) \Big] \\
S_{r, \kappa}^{C1\to S}[Q, Q'] &=&
\int_{t_{i}}^{t_{f}} \mathrm{~d} t \int_{t_{i}}^{t} \mathrm{~d} 
s\Big[ k_{r}^{C1\to S}(t-s) \big(Q(t) Q(s)+\nonumber \\ 
&& Q'(t) Q'(s)\big) -  k_{r}^{C1\to S}\left(t-s+\kappa \right) Q(t) Q'(s) \nonumber \\
&& \quad -\, k_{r}^{C1\to S}\left(t-s-\kappa\right) Q(s) Q'(t) \Big].
\end{eqnarray}
with the same kernels as in
(\ref{eq:S_i-B-to-C},\ref{eq:S_r-B-to-C}).
In the setting 
of the bath interacting with the cavity
the form of the new kernels 
is still that of a modification of 
(\ref{eq:S_i-B-to-C},\ref{eq:S_r-B-to-C})
when the coupling constants are time dependent, 
but the time shifts then only
pertain to the arguments of the sines and the cosines 
see {\it e.g.} \cite{AurellKawaiGoyal2020}. 

The first moment of the energy change
follows from the generating function 
by differentiating with respect to 
parameter $\nu$ and then setting it to zero,
which gives
\begin{eqnarray}
\label{eq:der-1}
&&\left.\frac{d}{d(i \kappa)} \mathcal{F}^{C1 \to S}_{\kappa}[Q, Q']\right|_{\kappa=0} =  \mathcal{F}^{C1\to S}[Q, Q'] \iint^{t_f,t} dt ds \nonumber\\ && \nonumber
\Bigg[ \left(Q(t) Q'(s)+ Q(s) Q'(t)\right) {\cal I}^{C1\to S}(t-s) \\
&&
\left(Q(t) Q'(s)- Q(s) Q'(t)\right) {\cal J}^{C1\to S}(t-s)   \Bigg]  \label{eq:der-2},
\end{eqnarray}
with two kernels
\begin{eqnarray}
{\cal I}^{C1 \to S}(t-s) &\equiv& \frac{1}{\hbar}\frac{d k_{i}^{C1\to S}\left(t-s\right)}{d(t-s)}  \\ 
{\cal J}^{C1 \to S}(t-s) &\equiv& -\frac{i}{\hbar}
\frac{d k_{r}^{C1\to S}\left(t-s\right) }{d(t-s)} 
\end{eqnarray}
It is convenient to extend these kernels,
analogous to the ones
introduced 
in \cite{AurellEichhorn2015}
for the situation where the system
interacts directly with the bath(s),
to negative time arguments. ${\cal I}^{C1 \to S}$
is then an even function and ${\cal J}^{C1 \to S}$ is odd.

\section{A Caldeira-Leggett bath seen through a cavity}
\label{sec:Caldeira-Leggett}
Many approaches have been developed
to deal with open quantum system
dynamics, numerically and analytically,
and several of them can be adapted 
to to quite complex interactions.
Here we focus instead on what is special
when inserting a single oscillator 
mode between a system and a real bath.
We therefore consider what is arguably
the simplest but still realistic and interesting
setting, which is a system interacting via
cavities with two Caldeira-Leggett baths.
In this section we will thus assume that
in both baths the oscillators form a continuous 
spectrum with
frequency-dependent bath-cavity interaction 
coefficients
$C_{\omega }$
and
spectral density $f(\omega )$ such that 
\begin{equation}
f(\omega )C_{\omega }^{2} =\left\{\begin{array}{ll} 
\frac{2\eta {{\omega }^{2}}}{\pi } & \hbox{if $\omega < \Omega$}\\
0 & \hbox{if $\omega > \Omega$} 
\end{array} \right.
\end{equation}
The parameter $\eta$ which can be the same or different for the
two baths has the dimension 
$\hbox{mass}\cdot \hbox{length}/\hbox{time}$
of a classical friction coefficient. 
When acting on processes on time scales
longer than $\Omega^{-1}$ the first Feynman-Vernon
kernel (from one of the baths) becomes ${k}_{i}^{B\to C} \approx -\eta \frac{{\rm d}}{{\rm d}(t-s )}\delta (t-s )$
and the bath-to-cavity action term $S_i^{B\to C}$ is a 
renormalization of the cavity potential plus a term $-\frac{\eta }{2}\int \Delta X \dot{\bar{X}}  dt$. 
The other kernel ${k}_{r}^{B\to C}$
is on the Fourier side 
$\eta \nu \,
\hbox{coth}\left(\frac{\nu\beta\hbar}{2}\right)$
which at frequencies small in scale of temperature ($\nu$ less than $(\hbar\beta)^{-1}$)
tends to $\frac{2\eta }{\beta\hbar}$.
In the time domain, for processes on scales slower
than $\hbar\beta$, ${k}_{r}^{B\to C}$
is a delta function,
and $S_r^{B\to C}$ tends to $\frac{\eta }{\beta\hbar}\int (\Delta X)^2 dt$.

The equation for the auxiliary function
$\Delta X^*$, in one of the cavities, is now that of a driven damped
oscillator
\begin{eqnarray}
\Delta \ddot{X}^* + \omega_C^2 \Delta X^* 
&=& C_{SC}(t)\Delta Q - \eta \Delta \dot{X}^*(s) \nonumber \\
\label{eq:Vernon-auxiliary-damped}
\Delta x^*_f=\Delta\dot{x}^*_f&=&0
\end{eqnarray}
For time-independent interactions 
that go on for all time, $\Delta X^*$
is hence a filtered version of $\Delta Q$
\textit{i.e.}
\begin{eqnarray}
\Delta \hat{X}^*(\nu) =
\frac{C_{SC} \Delta \hat{Q}(\nu)}
{-\nu^2+ \omega_C^2+i\eta\nu} 
\label{eq:Vernon-auxiliary-damped-Fourier}
\end{eqnarray}
The denominator is the response function in
the Fourier domain.
The simple formulae 
\eqref{eq:Vernon-auxiliary-damped-Fourier}
or
\eqref{eq:Vernon-response-damped-Fourier}
can be 
inserted directly in
\eqref{eq:Vernon-transform-Fourier-side}
and
\eqref{eq:Vernon-transform-Fourier-side-real}
to give
\begin{eqnarray}
\label{eq:real-Vernon-transform-kernel-0-nu-simple}
\hat{k}_i^{C1\to S}(\nu) &=&
\frac{1}{2}
\frac{ C_{SC}^2} 
{-\nu^2+ \omega_C^2 +i\eta\nu} \\
\hat{k}_r^{C1\to S}(\nu) &\approx&
\frac{\eta \nu}{2} 
\frac{ C_{SC}^2 \, \hbox{coth}\left(\frac{\nu\beta\hbar}{2}\right)}
{(-\nu^2+ \omega_C^2)^2+\eta^2\nu^2}
\end{eqnarray}
The approximation symbol in the second equation is to emphasize
that boundary terms which are in principle
also present, have been neglected.
Obviously both $\hat{k}_i^{C\to S}$
and $\hat{k}_r^{B\to C (S)}$ are largest
at frequencies $\nu\approx\omega_C$
and go down away from this resonance.
This means out of all the possible frequencies
of the motion of the system, it is only those around the cavity
frequency which feel the bath strongly, all the others 
only experience the bath indirectly.

We now turn to thermodynamics.
As shown in Section~\ref{eq:thermo}
the expected energy change 
in the cavity and the bath on side 1,
in a process that goes on for a long time, is
\begin{widetext}
\begin{eqnarray}
\left<\Delta E_{B1} + \Delta E_{C1}\right>_{if}  &=&  \int_{if} {\cal D}Q {\cal D} Q' e^{\frac{i}{\hbar}(S_S[Q]-S_S[Q'])+\frac{i}{\hbar}(S^{C1\to S}_i[Q,Q']+S^{C2\to S}_i[Q,Q'])
-\frac{1}{\hbar}(S^{C1\to S}_r[Q,Q']+S^{C2\to S}_r[Q,Q'])} \nonumber\\ && \iint^{t_f,t_f} dt ds 
\Bigg[ Q(t) Q'(s) \left({\cal I}^{C1\to S}(t-s)  + {\cal J}^{C1\to S}(t-s)\right) \Bigg] 
\label{eq:expected-energy-side-1},
\end{eqnarray}
\end{widetext}
where the two kernels in the second line are the derivatives of the 
of the Feynman-Vernon kernels on side 1 with respect to the time argument.
When the bath is as Caldeira-Leggett model, on the Fourier side
this means
\begin{eqnarray}
\hat{\cal I}^{C1 \to S}(\nu) &=&
i\frac{\nu}{2\hbar}
\frac{ C_{SC1}^2} 
{-\nu^2+ \omega_{C1}^2 +i\eta_1\nu}\\
\hat{\cal J}^{C1 \to S}(\nu) &=& 
\frac{\eta_1 \nu^2 }{2\hbar }
\frac{ C_{SC1}^2 \, \hbox{coth}\left(\frac{\nu\beta_1\hbar}{2}\right)}
{(-\nu^2+ \omega_{C1}^2)^2+\eta_1^2\nu^2}
\end{eqnarray}
The long first line in \eqref{eq:expected-energy-side-1} expresses a quantum expectation value
of the system process. This will depend on the nature of the system and its own
dynamics as much as how it interacts with the cavities and through them with the baths. 
However, if 
the system reaches a stationary state $\hat\rho^{Stat.}_S$, then we 
can take that as initial state at
$t_i'$
sufficiently before $t$ and $s$, and integrate
out the process from $t_i$ to $t_i'$.
There is then no memory of the initial
system state $i$.
Likewise we can integrate out the process from
the largest of $t$ and $s$ to $t_f$
such that at the end we only find the
probability to observe the system in state
$f$ at the final time,
$\langle f|\hat\rho^{Stat.}_S|f \rangle$.
The path integrals 
in
\eqref{eq:expected-energy-side-1} 
are hence averages in stationary state 
of the super-operators 
which the path integral variables $Q(t)$
and $Q'(s)$
represent: $Q(t)$ means acting with the operator
$\hat{Q}(t)$ from the left, and
$Q'(s)$ means acting with the operator
$\hat{Q}(s)$ from the right.
If the final state 
$f$ is summed over we can
write the averages as
an open system correlation function
\begin{equation}
C_{Q}(\tau)=
\hbox{Tr}\left[\left< \hat{Q}(\tau)\hat\rho^{Stat.}_S \hat{Q}(0)   \right>\right]
\label{eq:non-equal-time-correlations}
\end{equation}
where $\tau=t-s$, $\hat\rho^{Stat.}_S$ is inserted
long before $t$ and $s$, the process is evolved
inserting whichever 
comes first of
$\hat{Q}(0)$ and $\hat{Q}(\tau)$, and then the 
other operator is inserted and the
trace is taken at 
the later time.
By the cyclic property of the trace the later 
of the two operators can be moved over to the other 
side, that is
$\hbox{Tr}\left[\left< \hat{Q}(0) \hat{Q}(\tau)  \hat\rho^{Stat.}_S\right>\right]$
if $\tau\leq 0$,
and
$\hbox{Tr}\left[\left< \hat\rho^{Stat.}_S \hat{Q}(0) \hat{Q}(\tau)  \right>\right]$
if $\tau\geq 0$.

The analytical/numerical determination of 
quantities as in \eqref{eq:non-equal-time-correlations}
is not a trivial task even in
simple
open quantum systems, for recent investigations
using different methods, see~\textit{e.g.} \cite{IvanovBreuer2015,Ban2018,Ban2019}.
Theoretically such quantities seem to have been first considered
by Lindblad in \cite{Lindblad1979},
and more recently by several groups~\cite{Aharonov2009,Fedrizzi2011,Silva2017,Ringbauer2018,Costa2018,Jordan2018,Nowakowski2018,Zhang2020,Milz2020}.
The open systems aspect is hence not here the crucial one: the central problem
is give operational meaning to operators inserted
at different times
on the left and on the right of a density matrix.
One approach is to use pre- and post-selection,
following \cite{Aharonov1964},
and more recently~\cite{Silva2017}.
The proposal in~\cite{Jordan2018} is that auxiliary systems
can be coupled to the system at different times
after which
\eqref{eq:non-equal-time-correlations} can
be realized as joint measurements of the auxiliary
systems at the end of the process, and after tracing out
the system.

When the process goes on for a long time 
the expected energy change in the cavity-and-bath 
per unit time is quantum thermal power,
which can hence be written
\begin{eqnarray}
\Pi &=& \int C_{Q}(\tau)
\left({\cal I}^{C1\to S}(\tau)
+ {\cal J}^{C1\to S}(\tau)\right)
\, d\tau  \\
&=& \int \frac{1}{2}\left(C_{Q}(\tau)+C_{Q}(-\tau)\right)
{\cal I}^{C1\to S}(\tau)\, d\tau \nonumber \\
&&
\quad +\, \int \Theta(\tau)\left(C_{Q}(\tau)-C_{Q}(-\tau)\right) {\cal J}^{C1\to S}(\tau)
\, d\tau \nonumber
\end{eqnarray}
where we have used that ${\cal I}^{C1\to S}$ is even and ${\cal J}^{C1\to S}$ is odd
and rewritten the last line so that it looks as a response function.
If both environments are at the same temperature
the above simplifies considerably.
For the sums and differences we can set
\begin{eqnarray}
S(\tau) &=& \hbox{Tr}\left[\left< \hat\rho^{\beta}_S \, \{ \hat{Q}(\tau),\hat{Q}(0)\}\,  \right>\right]
\\
\chi(\tau) &=& \hbox{Tr}\left[\left< \hat\rho^{\beta}_S \, \Theta(\tau) \left[ \hat{Q}(\tau),\hat{Q}(0)\right]  \right>\right]
\end{eqnarray}
the Fourier transforms of which are related by the fluctuation-dissipation theorem. 
Adding and subtracting terme
thermal power is then zero.

\section{Discussion}
\label{sec:discussion}
In this work we have considered the problem of estimating
energy taken from or given to a system when it is connected 
to a bath through a cavity. The cavity is a harmonic oscillator
degree of freedom interacting linearly with both the 
bath and the system. This setting describes 
a superconducting qubit connected to two normal-metal baths
through transmission lines as investigated experimentally
and theoretically in~\cite{Ronzani2018} and other publications
from the same group.
We have considered the system connected to one bath
or to two baths, possibly at different temperatures.

The paper has two main points. The first is that the influence from
a bath on a cavity is transformed in a systematic manner to an influence from the
cavity to a bath. Although not a new result 
this is very useful. We have called this transformation
the \textit{Vernon transform}, as it first appeared in
an appendix to F.~Vernon's (unpublished) 1959 PhD thesis from Caltech.
On the Fourier side the transform is similar to a band-pass
filter, such that the system mostly experiences the influences
of the bath(s) at
the resonance frequencies of the cavities.
The form of this filter can be described in a precise manner
both for dissipation and for quantum noise.
We hope to have helped bring light again to this nice 
classical result.
The second point is that one can combine the Vernon transform 
with recent results in quantum thermodynamics to get
compact expressions for the generating function of quantum
heat (energy change in the environment).
To do so in a simple way one has to consider the bath and the cavity as one environment;
the moments of energy changes in the bath only can also be determined
by the same techniques,
but involve considerably more complicated expressions.

As an example we have considered thermal power (expected energy change in
the environment per unit time) when the bath is of the Caldeira-Leggett 
type. Thermal power is a convolution of a kernel describing the bath
given by the Vernon transform,
and an open system unequal-time correlation function. The latter is
not straight-forward to compute analytically -- there is no free lunch -- 
but could be determined experimentally using methods of modern
quantum science, or estimated from 
numerical solutions of the open system dynamics.
In any case, we hope to have shown that the problem of
heat transfer through a system connected to two baths
via two cavities admits a systematic theory for all 
strengths of the interactions.

\section*{Acknowledgments}
We thank Roberto Mulet for numerous
discussions.
This work
was supported by 
the Swedish Research Council grant 2020-04980 (E.A.),
and by the European Research
Council grant 742104 (J.T.).

\bibliography{references}

\appendix

\section{Vernon's derivation of Feynman-Vernon action}
\label{sec:vernon-feynman-vernon}
In this section we give for completeness
details omitted in Section~\ref{sec:Vernon}.
This derivation 
Feynman-Vernon action kernels can be found in
Appendix~1 of Vernon's PhD thesis~\cite{VernonPhD}, 
as well as elsewhere in the later literature \textit{e.g.}~\cite{CALDEIRA1983},
though not in~\cite{FeynmanVernon1963}.
We start by only considering the cavity and the bath.
The evolution operator corresponding
to $\hat{H}_C+\hat{H}_B + \hat{H}_{BC}(t)$
acting on cavity-bath wave function
can be written as a (multi-variable) path integral
\begin{eqnarray}
\hat{U}&=& \int {\cal D}X {\cal D}Y\,
e^{\frac{i}{\hbar}\left(S_C[X] + S_B[Y]+S_{CB}[X,Y]\right)} 
\end{eqnarray}
The evolution operator acting on cavity-bath density matrices
is similarly a double path integral over ``forward paths'' and a ``backward paths''
\begin{eqnarray}
\label{eq:quantum-map-pure}
\hat{U}\cdot \hat{U}^{\dagger} &=& \int {\cal D}X {\cal D}X' {\cal D}Y {\cal D}Y'  e^{\frac{i}{\hbar}\left(S_C[X]+S_B[Y]+S_{CB}[X,Y]\right)} \nonumber \\
&&\quad\cdot\quad 
e^{-\frac{i}{\hbar}\left(S_C[X'] +S_B[X'] +S_{CB}[X',Y']\right)} 
\end{eqnarray}
where the slot marks where the initial density matrix is to be inserted.
When the cavity and the bath are initially independent and the bath is
initially in a thermal state, the path integrals for each mode of the bath 
can be done independently. 
When the final state of a bath oscillator is traced over,
the outcome is the Feynman-Vernon influence functional
$\mathcal{F}^{k\to C}[X, X']$, where by
$k\to C$ we here mean the influence of bath 
oscillator $k$ on the cavity.
\\\\
Vernon's approach to compute $\mathcal{F}^{k\to C}[X, X']$
starts by re-writing the double path integral in terms of
of sums and differences of forward and backward paths 
\textit{i.e.} in terms of $\bar{Y}=Y+Y'$,$\Delta{Y}=Y-Y'$,
$\bar{X}=X+X'$ and $\Delta{X}=X-X'$.
The initial thermal state of bath oscillator $k$ 
is then $\rho_B^{(k)}(\bar{y},\Delta y)= \frac{1}{N}\exp\left(-\frac{\omega_k}{4\hbar}\left(
\bar{y}^2 \tanh\frac{\omega_k\beta\hbar}{2}+
\Delta{y}^2\hbox{coth}\frac{\omega_k\beta\hbar}{2}\right)\right)$,
where $N$ is a normalization.
The actions over the paths are for the terms 
that involve the bath variable
$$
\frac{1}{2} \int_{t_i}^{t_f} \dot{\bar{Y}} \Delta \dot{Y} - \omega_k^2 \bar{Y} \Delta Y 
+ C_{CB}(t)\left(\bar{X}\Delta Y + \Delta X\bar{Y}
\right)\, dt
$$
where the first term can be integrated by parts to
$-\frac{1}{2} \int_{t_i}^{t_f} \bar{Y} \Delta \ddot{Y} \, dt
+ \frac{1}{2} \left(\bar{y}_f \Delta \dot{y}_f -\bar{y} \Delta \dot{y}\right)$.
\\\\
Tracing over the final state of the bath oscillator means that
$\Delta y_f=\Delta Y(t_f)$ has to be zero, while $\bar{y}_f
=\bar{Y}(t_f)$ is integrated over. 
The terms linear in $\bar{Y}$ and $\bar{y}_f$ in the action over 
paths is then $\frac{1}{2} \bar{y}_f \Delta \dot{y}_f$ and
\begin{equation}
\label{eq:action-partial-int}
\frac{1}{2} \int_{t_i}^{t_f} \bar{Y}\left(-\Delta \ddot{Y} - \omega_k^2 \Delta Y 
+ C_{CB}(t)\Delta X\right)\, dt
\end{equation}
The initial condition and the 
integration over $\bar{y}_f$ and $\bar{Y}$
means that path integral is non-zero only when $\Delta Y$
agrees with a deterministic auxiliary function 
defined in the main text, and for convenience repeated here
$$\Delta y^*_f=\Delta\dot{y}^*_f=0\quad \Delta \ddot{Y}^* = - \omega_k^2 \Delta Y^* 
+ C_{CB}(t)\Delta X=0 $$
As discussed in main text this
is the equation of a harmonic oscillator 
with an external drive starting from rest at $t_f$
and the solution can be expressed as
\begin{equation}
\label{eq:response-appendix}
\Delta Y^*(t) = \int_t^{t_f} R_k(t,s)  C_{CB}(s) \Delta X(s) ds
\end{equation}
where $R_k$ is a response function.
\\\\
On the other hand, \eqref{eq:action-partial-int}
can be seen as 
$\int_{t_i}^{t_f} \bar{Y}\left(L \Delta Y \right)   dt$
where $L$ is a linear operator acting on $\Delta Y$,
hence the functional integral also formally gives a factor
$\frac{1}{|\det L|}$. This factor is the same as appears as
$\frac{1}{\sqrt{|\det L|}}$ 
in the ordinary path integral of the harmonic oscillator,
and should therefore be interpreted as undetermined constant 
times $\sin^{-1}\omega_k(t_f-t_i)$.
The last integral over $\bar{y}$ yields 
$\exp\left(-\frac{1}{4\omega_k\hbar} \left(\Delta\dot{y}\right)^2\hbox{coth}\frac{\omega_k\beta\hbar}{2}\right)$
(cancelling the normalization factor $N$ in above).
What remains of the actions involving 
the sum of forward and backward paths ($\bar{y}_f$, $\bar{Y}$ and $\bar{y}$)
is hence this term in $\left(\Delta\dot{y}\right)^2$, the other remaining term 
from $\rho_B^{(k)}$ which is 
$\exp\left(-\frac{\omega_k}{4\hbar}\left(
\left(\Delta{y}\right)^2\hbox{coth}\frac{\omega_k\beta\hbar}{2}\right)\right)$,
and
\begin{eqnarray}
\label{eq:final-value}
&& \hbox{Const.}\cdot \frac{1}{|\det L|}\cdot e^{\frac{i}{2\hbar}\int_{t_i}^{t_f} 
C_{CB}(t)\bar{X}\Delta Y\, dt} \cdot \nonumber \\
&&\quad
\delta(\Delta y_f - \Delta y^*_f) \delta(\Delta \dot{y}_f - \Delta \dot{y}^*_f) \delta(\Delta Y-\Delta Y^*)
\end{eqnarray}
We can trade the integral over $\Delta y$ against an integral over 
$\Delta \dot{y}_f$ at the price of Jacobian which cancels
$\frac{1}{|\det L|}$.
Hence the Feynman-Vernon action from the bath on the cavity can
be expressed as 
\begin{eqnarray}
\label{eq:Feynman-Vernon-bath-cavity-appendix}
\mathcal{F}^{k\to C} &=&
e^{\frac{i}{\hbar}\int_{t_i}^{t_f} 
\frac{C_{CB}(t)}{2}\bar{X}\Delta Y^*\, dt} \nonumber \\
&& e^{-\frac{1}{4\hbar}
\hbox{coth}\frac{\omega_k\beta\hbar}{2}
\left(\frac{1}{\omega_k} \left(\Delta\dot{y}^*\right)^2
+ \omega_k\left(\Delta{y}^*\right)^2\right)}
\end{eqnarray}
The imaginary term in the exponent in above (real part of Feynman-Vernon
action) is discussed in the main text.
Here we will continue on the 
real terms in the exponent (imaginary part of Feynman-Vernon
action).

Using \eqref{eq:response-appendix} and 
 $R_k=\frac{1}{\omega_k}\sin\omega_k(s-t)$ we have
\begin{eqnarray}
\Delta y^* &=& \int_{t_i}^{t_f} \frac{1}{\omega_k}\sin\omega_k(t-t_i) C_{CB}(t) \Delta X(t)\, dt
\end{eqnarray}
and hence 
\begin{eqnarray}
&&\left(\frac{1}{\omega_k^2}\Delta\dot{y}^*\right)^2
+ \left(\Delta{y}^*\right)^2  
= \nonumber \\
&&\iint^{t_f}_{t_i} \Delta X(t)\Delta X(s) \big(\cos\omega_k(t-t_i)\cos\omega_k(s-t_i) + \nonumber \\
&&\sin\omega_k(t-t_i)\sin\omega_k(s-t_i)\big)C_{CB}(t)C_{CB}(s)  \, dt \, ds
\end{eqnarray}
where the inner parenthesis is $\cos\omega_k(t-s)$.
This hence gives the imaginary part of the Feynman-Vernon
action as in the main text \eqref{eq:S_r-B-to-C}.

\section{The real part of the Vernon transform}
\label{sec:vernon-real}
In this appendix we show the real parts
of the Vernon transform, found in
Appendix~5 of Vernon's PhD thesis~\cite{VernonPhD}.

From section~\ref{sec:Vernon} we know how the Vernon transform 
${\cal V}$ transforms the Feynman-Vernon kernel
$k_i^{B\to C}$  on the
Feynman-Vernon kernel
$k_i^{C\to S}$.
Collecting the various terms of that derivation we
will now describe the real 
Vernon transform ${\cal W}$ 
which maps
$k_r^{B\to C}$  on
$k_r^{C\to S}$. 
This will depend on 
on $k_i^{C\to S}$, hence  ${\cal W}$ depends on ${\cal V}$.

The first term to consider is the imaginary part of the
Feynman-Vernon
action of the bath on the cavity, given as \eqref{eq:S_r-B-to-C}
in the main text, but expressing the cavity variable in the 
system variable through \eqref{eq:response-2}.
This will be
\begin{eqnarray}
\label{eq:S_r-B-to-C-in-terms-of-S}
S^{C\to C}_r&=&  -\frac{1}{2\hbar}
\iint^{t_f} k_r^{B\to C}(t,s) 
\big(\int_t^{t_f} R_C(t,t') \nonumber \\ 
&& \qquad C_{SC}(t') \Delta Q(t') \cdot \int_s^{t_f} R_C(s,s') \nonumber \\ 
&& \quad C_{SC}(s') \Delta Q(s')\, \big) \, ds\,  dt 
\end{eqnarray}
which can be re-written as
\begin{eqnarray}
\label{eq:S_r-B-to-C-in-terms-of-S-2}
S^{C\to S}_r&=&  -\frac{1}{2\hbar}
\iint^{t_f} k_r^{C\to S}(t',s') \nonumber \\
&&\qquad \Delta Q(t') \Delta Q(s') ds'\,  dt' 
\end{eqnarray}
with the new combined kernel
\begin{eqnarray}
\label{eq:real-Vernon-transform-kernel-0}
k_r^{C\to S}(t',s') &=& \int_{t_i}^{s'}\int_{t_i}^{t'} k_r^{B\to C}(t,s) C_{SC}(t') C_{SC}(s') \nonumber \\
&& R_C(t,t') R_C(s,s')\, ds\, dt
\end{eqnarray}
The response functions can then further be expressed in terms the kernel $k_i^{C\to S}$ through
\eqref{eq:Vernon-transform-def} so that we have also
\begin{eqnarray}
k_r^{C\to S}(t',s') &=& \int_{t_i}^{s'}\int_{t_i}^{t'}  \frac{k_i^{C\to S}(t,t') k_i^{C\to S}(s,s')}{C_{SC}(t) C_{SC}(s)}\nonumber \\
&& k_r^{B\to C}(t,s)    \, ds\, dt
\end{eqnarray}
Note that $k_i^{C\to S}$ depends on two interaction coefficients, hence the fraction in above is in total quadratic in
in the cavity-system interaction.
The above is in fact the only term proportional
to the total duration of the process, and the only
one considered in the main body of the paper.

The second term to consider is a boundary term,
the analogy of the real exponent in
\eqref{eq:Feynman-Vernon-bath-cavity-appendix} but for the cavity.
This will give
\begin{eqnarray}
\label{eq:S_r-C-to-S-from-C}
S^{C\to S (S)}_r&=& -\frac{1}{4\hbar}
\hbox{coth}\frac{\omega_C\beta\hbar}{2}
\left(\frac{1}{\omega_C}\left(\Delta\dot{x}^*\right)^2
+ \omega_C\left(\Delta{x}^*\right)^2\right) \nonumber
\end{eqnarray}
where again the cavity variable should be expressed in terms of the 
system variable through \eqref{eq:response-2}.
We write this as
\begin{eqnarray}
\label{eq:response-2-app-B}
\Delta x^* &=& \int_{t_i}^{t_f} R_C(t_i,t) C_{SC}(t) \Delta Q(t) dt \\
\Delta \dot{x}^* &=& \int_{t_i}^{t_f} \dot{R}_C(t_i,t) C_{SC}(t) \Delta Q(t) dt 
\end{eqnarray}
where $\dot{R}_C(t_i,t) = \frac{d}{ds}R_C(s,t)|_{s=t_i}$.
This then gives 
\begin{eqnarray}
\label{eq:S_r-C-to-S-from-C-2}
S^{C\to S (S)}_r&=&  -\frac{1}{2\hbar}
\iint^{t_f} \left(k_r^{C\to S (S,1)} + k_r^{C\to S (S,2)}\right)\nonumber \\
&&\qquad \Delta Q(t') \Delta Q(s') ds'\,  dt' 
\end{eqnarray}
with two kernels
\begin{eqnarray}
\label{eq:real-Vernon-transform-kernel-1}
k_r^{C\to S (S,1)}(t',s') &=& 
\frac{\omega_C}{2} \hbox{coth}\frac{\omega_C\beta\hbar}{2} C_{SC}(t') C_{SC}(s')  \nonumber \\
&&\qquad R_C(t_i,t') R_C(t_i,s') \\
\label{eq:real-Vernon-transform-kernel-2}
k_r^{C\to S (S,2)}(t',s') &=& \frac{1}{2\omega_C} \hbox{coth}\frac{\omega_C\beta\hbar}{2} C_{SC}(t') C_{SC}(s')  \nonumber \\
&&\qquad \dot{R}_C(t_i,t') \dot{R}_C(t_i,s')
\end{eqnarray}
We summarize this as
\begin{defi} \label{def:Vernon-transform-real-detailed}
The real Vernon transform ${\cal W}$
is a linear mapping of the
Feynman-Vernon kernel
$k_r^{B\to C}$  on the
Feynman-Vernon kernel
$k_r^{C\to S}$ given by the
the integral transform
\eqref{eq:real-Vernon-transform-kernel-0}
which is proportional to $k_r^{B\to C}$,
and 
two boundary terms
\eqref{eq:real-Vernon-transform-kernel-1} and
\eqref{eq:real-Vernon-transform-kernel-2}
which do not depend on $k_r^{B\to C}$.
\end{defi}
\noindent
We end this appendix by discussing simplifications
of
\eqref{eq:real-Vernon-transform-kernel-0}.
If the interaction is time-independent 
for all time
and $k_r^{B\to C}$ and the response function $R_C$ only depend on the 
time differences, 
$k_r^{C\to S}$
will also only depend on the time difference, and we have
\begin{eqnarray}
\label{eq:real-Vernon-transform-kernel-0-simple}
k_r^{C\to S}(\tau) &=& \iint^{0,\tau}  
C_{SC}^2 \, k_r^{B\to C}(t-s) \nonumber \\
&& R_C(-t) R_C(\tau-s)\, ds\, dt
\end{eqnarray}
On the Fourier side that gives
\begin{eqnarray}
S^{B\to C (S)}_r&=&  -\frac{1}{2\hbar}
\int\hat{k}_r^{B\to C (S)}(\nu) \nonumber \\
&&\qquad \Delta \hat{Q}(\nu) \Delta \hat{Q}(-\nu) d\nu 
\end{eqnarray}
with
\begin{eqnarray}
\label{eq:real-Vernon-transform-kernel-0-nu}
\hat{k}_r^{B\to C (S)}(\nu) &=&
\hat{R}_C(\nu) \hat{R}_C(-\nu)
\hat{k}_r^{B\to C}(-\nu)
\end{eqnarray}
In other words, on the Fourier side this transformed
imaginary-action Feynman-Vernon kernel from the cavity to the
system is proportional to the one from the bath to the cavity
at minus the frequency, the proportionality being the spectral energy 
of the response function.

The two terms 
\eqref{eq:real-Vernon-transform-kernel-1} and
\eqref{eq:real-Vernon-transform-kernel-2}
stem from the parenthesis in \eqref{eq:S_r-C-to-S-from-C}
which is like a final energy for the auxiliary process.
If there is no damping this energy could grow indefinitely,
but with damping it will remain finite.
These two terms hence gives a contribution which does
not depend on the duration of the process, and can hence be
ignored when considering quantities per unit time such as quantum power.

\section{Vernon transforms for the energy changes in a bath alone}
\label{app:vernon-transform-heat-bath-alone}
We will show how the Vernon transform of Section ~\ref{sec:Vernon} can be used to compute moments of generating functions of energy changes in the bath, in this context also called quantum heat.
For simplicity we focus on the case in which the system is coupled via the cavity to only one bath.
Generating functions for energy changes in multiple baths
have been considered several times, \textit{e.g.} 
recently in \cite{AurellDonvilMallick2020}.
For the simpler one-sided case it reads
\begin{eqnarray}
G_{i f}(\nu)&=&\operatorname{Tr}_{CB}
\langle f| e^{i \nu H_{B}}U(t) e^{-i \nu H_{B}} \nonumber \\&& \times \left(\rho_{C} \otimes \rho_{\beta_{B}} \otimes|i\rangle\langle i| \right) U^{\dagger}(t) |f\rangle, 
\label{eq:Gif-nu}
\end{eqnarray}
where $|i\rangle$ ($|f\rangle$) denotes the initial (final) state of the system, $\rho_{C}$ and $\rho_{\beta_{B}}$ are initial states of the cavity and the bath (one each), respectively, and $U(t)$ is the evolution of the system, cavity, and the bath generated by the Hamiltonian 
\eqref{eq:total-Hamiltonian}.

Moments of the energy change are generated by taking derivatives of the generating function with respect to $\nu$, e.g. the change of average energy of the bath is given by the first derivative  
\begin{eqnarray}
\label{eq:expected-quantum-heat}
\left\langle\Delta E_{B}\right\rangle=\left.\frac{d}{d(i \nu)} G_{i f}(\nu)\right|_{\nu=0}.
\end{eqnarray}
The path integral formulation of generating functions and moments
of heat and work has been studied in 
\cite{AurellKawaiGoyal2020}.

The most important change with respect to the setting of Section~\ref{sec:Vernon} is that the path integral formulation of the generating function involves modified Feynman-Vernon kernels $k_i^{B \to C}, k_r^{B \to C}$. Following the notation of the previous Section one has 

\begin{eqnarray}
G_{if}(\nu)&=&\int_{if}  {\cal D} Q {\cal D} Q' 
   e^{\frac{i}{\hbar} \int_{t_i}^{t_f} dt \left( S_S[Q]-S_S[Q']
   \right)} \rho_S\left(Q,Q' \right) \nonumber \\ 
   && \quad \int  {\cal D} X {\cal D} X'    
e^{\frac{i}{\hbar}\left( S_{CS}[Q,X]-S_{CS}[Q',X']\right)}  
   \nonumber \\
&& \qquad e^{\frac{i}{\hbar}\left(
S_{C}[X]-S_{C}[X']
   \right)}  
   \mathcal{F}^{B \to C}_{\nu}[X, X']  \rho_C\left(X,X' \right),  
   \label{eq:G-nu-def}
\end{eqnarray}
where the only change in 
is the substitution of
the Feynman-Vernon functional
$\mathcal{F}^{B\to C }[X, X']$
with
\begin{eqnarray}
    \begin{aligned}
\mathcal{F}^{B\to C}_{\nu}[X, X'] &=e^{\frac{i}{\hbar}S_{i, \nu}^{B}[X, X']-\frac{1}{\hbar} S_{r, \nu}^{B}[X, X']}
\end{aligned}
\end{eqnarray}
The imaginary and real part of this modified Feynman-Vernon action depend on $\nu$. 
For the case when the bath-cavity coupling
coefficients are constant they read
\begin{eqnarray}
S_{i, \nu}^{B}[X, X'] &=& \int_{t_{i}}^{t_{f}} \mathrm{~d} t \int_{t_{i}}^{t} \mathrm{~d} s\Big[ k_{i}^{B}(t-s) \big(X(t) X(s) \nonumber \\
&& -X'(t) X'(s)\big)  +   k_{i}^{B}\left(t-s+\nu\right) X(t) X'(s)
\nonumber \\
&& \quad -\,k_{i}^{B}\left(t-s-\nu\right)X(s) X'(t) \Big] \\
S_{r, \nu}^{B}[X, X'] &=&
\int_{t_{i}}^{t_{f}} \mathrm{~d} t \int_{t_{i}}^{t} \mathrm{~d} 
s\Big[ k_{r}^{B}(t-s) \big(X(t) X(s)+\nonumber \\ 
&& X'(t) X'(s)\big) -  k_{r}^{B}\left(t-s+\nu \right) X(t) X'(s) \nonumber \\
&& \quad -\, k_{r}^{B}\left(t-s-\nu\right) X(s) X'(t) \Big].
\end{eqnarray}
with the same kernels as in
(\ref{eq:S_i-B-to-C},\ref{eq:S_r-B-to-C}).
When the coupling constants are time dependent, 
the form of the new kernels is still that of a modification of 
(\ref{eq:S_i-B-to-C},\ref{eq:S_r-B-to-C}) but with time shifts
only to the arguments of the sine and the cosine 
respectively; 
see {\it e.g.} \cite{AurellKawaiGoyal2020}. 

The task is now to integrate out the cavity degrees of
freedom with the modified Feynman-Vernon actions from
the bath on the cavity.
This cannot be done with 
the Vernon transform
directly, because in terms of the sums and differences
of the forward and backward cavity paths
($\bar{X}$ and $\Delta X$), the modified
Feynman-Vernon action is quadratic in $\bar{X}$.
This is in contrast to the situation in Section~\ref{sec:Vernon}, where it was only linear in $\bar{X}$, and where the path 
integral therefore gave a functional delta.
Although the full generation function 
could also be addressed,
here we will focus on the moments of the
energy changes where this issue does not appear.

For the first moment, computed by the
formula \eqref{eq:expected-quantum-heat}
and expressed separately in $\bar{X}$ and $\Delta X$
we have three terms
\begin{eqnarray}
\label{eq:der-1}
&&\left.\frac{d}{d(i \nu)} \mathcal{F}^{C \to S}_{\nu}[\bar{Q}, \Delta Q]\right|_{\nu=0} =  \int {\cal D}\bar{X} {\cal D} \Delta X \iint^{t_f,t} dt ds \nonumber\\ && \nonumber
\Bigg[ \Delta X(t) \Delta X(s) {\cal I}^{B\to C}(t,s)    + \bar{X}(t) \Delta X(s) {\cal J}^{B\to C}(t,s) + \label{eq:der-1} \\&& \bar{X}(t) \bar{X}(s){\cal I}^{B\to C}(t,s)   \Bigg]  \mathcal{F}^{B\to C}_{\nu=0}[\bar{X}, \Delta{X}] \rho_C \left(\bar{x}_i,\Delta x_i \right) \label{eq:der-2},
\end{eqnarray}
with two kernels
\begin{eqnarray}
{\cal I}^{B \to C}(t,s) &\equiv& \frac{1}{4}\sum_k  C_{CB}(t)
C_{CB}(s) \nonumber \\ 
&& \cos \left[\omega_k(t-s)\right] \\
{\cal J}^{B \to C}(t,s) &\equiv& \frac{i}{2}\sum_k C_{CB}(t) 
C_{CB}(s) \nonumber \\ 
&&\qquad \sin \left[\omega_k(t-s)\right]  
 \coth \frac{\omega_k \beta}{2}
\end{eqnarray}
In above we have for completeness reintroduced the 
time dependence of the interaction coefficients.
The modified Feynman-Vernon kernels ${\cal I}$ 
and ${\cal J}$ were
derived in \cite{AurellEichhorn2015}.

The $\Delta X$ terms in \eqref{eq:der-1}
will by the path integral over $\bar{X}$
be transformed into the auxiliary function
$\Delta X^*$ which by \eqref{eq:Vernon-auxiliary-2}
and \eqref{eq:response-2} is a linear functional
of the system difference $\Delta Q$.
The $\bar{X}$ terms in \eqref{eq:der-1}
can similarly be expressed as functional
derivatives at zero with respect to a new
auxiliary field $L(t)$,
by adding a new coupling term 
$\frac{1}{2} \int^{t_f}_{t_i} dt' L(t') \bar{X}(t')$
to the action.
The path integral with respect to $\bar{X}$
then gives a new auxiliary function
$\Delta X_L^*$ which is a linear functional
of both $\Delta Q$ and $L$.

The structure of the expected heat functional in terms
of the system variables will hence be
\begin{eqnarray}
\frac{d}{d (i\nu)}G_{if}(\nu)|_{\nu=0}&=&\int_{if}  {\cal D} Q {\cal D} Q' 
   e^{\frac{i}{\hbar} \int_{t_i}^{t_f} dt \left( S_S[Q]-S_S[Q']
   \right)} \nonumber \\
\quad\cdot\quad \mathcal{F}^{C \to S}[\bar{Q}, \Delta Q] &\cdot& 
\iint^{t_f,t_f} \Big[ \Delta Q(t) \Delta Q(s) {\cal L}^{C\to S}_{\Delta Q, \Delta Q}(t,s) \nonumber \\
&& \quad +\quad \bar{Q}(t) \Delta Q(s) {\cal L}^{C\to S}_{\bar Q, \Delta Q}(t,s) \quad + \nonumber \\
&& \quad \bar{Q}(t) \bar{Q}(s) {\cal \bar{I}}^{C\to S}_{\bar Q, \bar Q}(t,s)   \Big]
\, ds\, dt
   \label{eq:I-and-J-for-C-on-S}
\end{eqnarray}
The only qualitative difference to what the
situation would be if the system would only
interact with the cavity 
(which then would take the role of the bath)
are that the kernels multiplying
$\Delta Q(t) \Delta Q(s)$, $\bar Q(t) \Delta Q(s)$ 
and $\bar{Q}(t) \bar{Q}(s)$, which are
\begin{eqnarray}
&&{\cal L}^{C\to S}_{\Delta Q, \Delta Q}(t,s)= \\&& \nonumber {\cal I}^{C\to S}_{\Delta Q, \Delta Q}(t,s) + {\cal J}^{C\to S}_{\Delta Q, \Delta Q}(t,s) + {\cal \bar{I}}^{C\to S}_{\Delta Q, \Delta Q}(t,s),
\end{eqnarray}
where the kernels $ {\cal I}^{C\to S}_{\Delta Q, \Delta Q}$, ${\cal J}^{C\to S}_{\Delta Q, \Delta Q}$, and ${\cal \bar{I}}^{C\to S}_{\Delta Q, \Delta Q}$ are given by Eqs (\ref{eq:idqdq}), (\ref{eq:jdqdq}), and (\ref{eq:bidqdq}). The second kernel 
\begin{eqnarray}
&&{\cal L}^{C\to S}_{\bar Q, \Delta Q}(t,s)= {\cal J}^{C\to S}_{\bar Q, \Delta Q}(t,s) + {\cal \bar{I}}^{C\to S}_{\bar Q, \Delta Q}(t,s),
\end{eqnarray}
where the kernels $ {\cal J}^{C\to S}_{\bar Q, \Delta Q}$, and  ${\cal \bar{I}}^{C\to S}_{\bar Q, \Delta Q}$ are given by Eqs (\ref{eq:jbqdq}), and (\ref{eq:bibqdq}). 

The last kernel ${\cal \bar{I}}^{C\to S}_{\bar Q, \bar Q}(t,s)$ is given by Eq. (\ref{eq:bibqbq})

It is clear that in principle the above
procedure can be also used to compute the higher moments of the generating function. It is also clear that 
the resulting expressions will be increasingly
more complicated convolutions involving 
bath-to-cavity quantities and response functions.

\subsection{Calculation details}
\label{app:heat-to-bath-alone}
Here we sketch the most important steps in performing the path integral over the cavity degree of freedom in  Eq. (\ref{eq:der-1}). As that expression consist of three terms we will describe the computation term by term. The first term reads 
\begin{eqnarray}
&&\int  {\cal D}  \bar{X} {\cal D} \Delta X  \iint^{t_f,t_f} dt ds {\cal I}^{B \to C}(t,s) \Delta X(t) \Delta X(s) \\&& \nonumber e^{\frac{i}{\hbar}  S_C\left[ \bar{X}, \Delta X \right]+\frac{i}{2 \hbar} \int dt C_{SC}(t) \left( \bar{Q}(t) \Delta X(t) + \Delta{Q}(t) \bar{X} \right) } \times \\&& \nonumber e^{\frac{i}{\hbar} \iint^{t_f,t_f} dt ds \left( \bar{X}(t) k_i^{B \to C}    (t-s) \Delta X(s) +i \Delta X(t) k_r^{B \to C}    (t-s) \Delta X(s)  \right) } \\&& \nonumber  \rho_C\left( \bar{X},\Delta X \right) \nonumber
\end{eqnarray}
The path integrals over $\Delta X, \; \bar{X}$ can be easily done 
\begin{eqnarray}
&&  \iint^{t_f,t_f} dt ds {\cal I}^{B \to C}(t,s) \Delta X^*(t) \Delta X^*(s) \\&& \nonumber e^{\frac{i}{2 \hbar} \int dt C_{SC}(t) \bar{Q}(t) \Delta X^*(t) -\frac{1}{\hbar}\int \int dt ds  \Delta X^*(t) k_r^{B \to C}    (t-s) \Delta X^*(s)   },
\end{eqnarray}
what translates into
\begin{eqnarray}
&&\iint^{t_f,t_f} dt ds \Delta Q(t) \Delta Q(s) {\cal I}^{C \to S}_{\Delta Q, \Delta Q}(t,s) \\&& e^{\frac{i}{2 \hbar} \int^{t_f}_{t_i} dt C_{SC}(t) \bar{Q}(t) \Delta X^*(t)-\frac{1}{\hbar}\int \int dt ds  \Delta X^*(t) k_r^{B \to C}    (t-s) \Delta X^*(s)}, \nonumber
\end{eqnarray}
where 
\begin{eqnarray}
\label{eq:idqdq}
{\cal I}^{C \to S}_{\Delta Q, \Delta Q}(t,s) = &&\iint^{t_f,t_f} dp dr C_{SC}(r) C_{SC}(p) R_C(t,r)R_C(s,p) \times \nonumber \\&& {\cal I}^{B \to C}(t,s). 
\end{eqnarray}
Now we move to the second term, which is
\begin{eqnarray}
&&\int {\cal D}  \bar{X} {\cal D} \Delta X \iint^{t_f,t_f} dt ds {\cal J}^{B \to C}(t,s) \bar X(t) \Delta X(s) \\&& \nonumber e^{\frac{i}{\hbar}  S_C\left[ \bar{X}, \Delta X \right]+\frac{i}{2 \hbar} \int dt C_{SC}(t) \left( \bar{Q}(t) \Delta X(t) + \Delta{Q}(t) \bar{X} \right) } \times \\&& e^{\frac{i}{\hbar}\iint^{t_f,t_f}  dt ds \left( \bar{X}(t) k_i^{B \to C}    (t-s) \Delta X(s) +i \Delta X(t) k_r^{B \to C}    (t-s) \Delta X(s)  \right) }. \nonumber 
\end{eqnarray}
We rewrite it using a functional derivative with respect to an auxiliary driving force $L(t)$ as 
\begin{eqnarray}
&&\frac{\hbar}{i}\int {\cal D}  \bar{X} {\cal D} \Delta X \iint^{t_f,t_f} dt ds {\cal J}^{B \to C}(t,s) \frac{\delta}{\delta L(t)} \Delta X(s) \\&& e^{\frac{i}{\hbar}  S_C\left[ \bar{X}, \Delta X \right ]+ \frac{i}{2 \hbar} \int dt C_{SC}(t) \bar{Q}(t) \Delta X(t) + \left( C_{SC}(t) \Delta{Q}(t) +L(t) \right) \bar{X}(t)  } \times \nonumber \\&& \nonumber e^{\frac{i}{\hbar} \iint^{t_f,t_f} dt ds \left( \bar{X}(t) k_i^{B \to C}    (t-s) \Delta X(s) +i \Delta X(t) k_r^{B \to C}    (t-s) \Delta X(s)  \right) }.
\end{eqnarray}
Now we can perform the integral over cavity degrees of freedom 
\begin{eqnarray}
&&\frac{\hbar}{i} \iint^{t_f,t_f} dt ds {\cal J}^{B \to C}(t,s)\frac{\delta}{\delta L(t)} \Delta X_L^*(s) \\&& \nonumber e^{\frac{i}{\hbar} S_S\left[ \bar{Q}, \Delta Q \right]+\frac{i}{2 \hbar} \int dt C_{SC}(t) \bar{Q}(t) \Delta X_L^*(t) } \times \\&& e^{-\frac{1}{\hbar} \iint^{t_f,t_f} dt ds  \Delta X_L^*(t) k_r^{B \to C}    (t-s) \Delta X_L^*(s)   }, \nonumber
\end{eqnarray}
Now we need to take the functional derivative with respect to the auxiliary field. It will result in three terms. The first contribution comes from the term multiplying the exponent
\begin{eqnarray}
\frac{\delta \Delta X^*(s)}{\delta L(t)} = R_C(s,t).
\end{eqnarray}
We are going to neglect this term.
The second term comes from the derivative of coupling term between the cavity and the system
\begin{eqnarray}
&&\frac{\delta }{\delta L(t)} \int dp C_{SC}(p) \bar{Q}(p) \Delta X_L^*(p) = \int dp C_{SC}(p) R_C(p,t)  \bar{Q}(p) \nonumber \\ 
\end{eqnarray}
We can rewrite it as
\begin{eqnarray}
&&\iint^{t_f} dt ds \bar Q(t) \Delta Q(s) {\cal J}^{C \to S}_{\bar Q, \Delta Q}(t,s) (t,s) \\&&e^{\frac{i}{2 \hbar} \int dt C_{SC}(t) \bar{Q}(t) \Delta X^*(t) -\frac{1}{\hbar}\int \int dt ds  \Delta X^*(t) k_r^{B \to C}    (t-s) \Delta X^*(s)} \nonumber 
\end{eqnarray}
where
\begin{eqnarray}
\label{eq:jbqdq}
&&{\cal J}^{C \to S}_{\bar Q, \Delta Q}(t,s) =  \\ \nonumber &&\frac{1}{2} C_{SC}(t) C_{SC}(s)\iint^{t_f} dp dr R_C(p,s)R_C(t,r) {\cal J}^{B \to C}(r,p) \end{eqnarray}
Moreover, the derivative of the term involving the real kernel will result in
\begin{eqnarray}
&&\iint^{t_f} dt ds \Delta Q(t) \Delta Q(s) {\cal J}_{\Delta Q, \Delta Q}(t,s) ^{C \to S}(t,s) \\&&e^{\frac{i}{2 \hbar} \int dt C_{SC}(t) \bar{Q}(t) \Delta X^*(t) -\frac{1}{\hbar}\int \int dt ds  \Delta X^*(t) k_r^{B \to C}    (t-s) \Delta X^*(s)} \nonumber
\nonumber \nonumber,
\end{eqnarray}
where
\begin{eqnarray}
\label{eq:jdqdq}
&&{\cal J}_{\Delta Q, \Delta Q}(t,s)=2i C_{SC}(t)C_{SC}(s) \iiiint^{t_f} dp dr du dw \nonumber \\ &&R_C(p,t) R_C(r,u)R_C(s,w)      k^{B \to C}_r(p-r)  {\cal J}^{B \to C}(u,w) \nonumber \\  
\end{eqnarray}
The third term involves the double functional derivative.
\begin{eqnarray}
&&\iint^{t_f,t_f} dt ds {\cal I}^{B \to C}(t,s)   \frac{\delta^2}{\delta L(t)\delta L(s)}  \\&& e^{\frac{i}{2 \hbar} \int dt C_{SC}(t) \bar{Q}(t) \Delta X_L^*(t) -\frac{1}{\hbar}\iint dt ds  \Delta X_L^*(t) k_r^{B \to C}    (t-s) \Delta X_L^*(s)   } \nonumber ,
\end{eqnarray}
It will result in three contributions. The first one is 
\begin{eqnarray}
&&\iint^{t_f} dt ds \bar Q(t) \bar{Q}(s) {\cal  \bar{I}}_{\bar Q, \bar Q}^{C \to S}(t,s) \\&&   e^{\frac{i}{2 \hbar} \int dt C_{SC}(t) \bar{Q}(t) \Delta X^*(t) -\frac{1}{\hbar}\int \int dt ds  \Delta X^*(t) k_r^{B \to C}    (t-s) \Delta X^*(s)} \nonumber
\end{eqnarray}
where 
\begin{eqnarray}
\label{eq:bibqbq}
&&{\cal I}_{\bar Q, \bar Q}^{C \to S}(t,s) = \\  &&\frac{1}{4}C_{SC}(t) C_{SC}(s) \iint^{t_f} dp dr R_C(p,t)R_C(r,s) {\cal I}^{B \to C}(p,t) \nonumber
\end{eqnarray}
The second one stems from the first derivative of the coupling term and the the real kernel term 
\begin{eqnarray}
&&\iint^{t_f} dt ds \bar Q(t) \Delta Q(s) {\cal  \bar{I}}_{\bar Q, \Delta Q}^{C \to S}(t,s) \cal \\&&e^{\frac{i}{2 \hbar} \int dt C_{SC}(t) \bar{Q}(t) \Delta X^*(t) -\frac{1}{\hbar}\int \int dt ds  \Delta X^*(t) k_r^{B \to C}    (t-s) \Delta X^*(s)} \nonumber
\nonumber \nonumber,,
\end{eqnarray}
where
\begin{eqnarray}
\label{eq:bibqdq}
&&{\cal  \bar{I}}_{\bar Q, \Delta Q}^{C \to S}(t,s) =\\
&&i C_{SC}(t) C_{SC}(s) \iiiint^{t_f} dp dr du dw  R_C(u,t)  R_C(p,s) R_C(r,w) \nonumber \\ && k^{B \to C}_r(p-r)  {\cal I}^{B \to C}(u,w)  \nonumber
\end{eqnarray}
and the second derivative with respect to the real kernel
\begin{eqnarray}
&&\iint^{t_f} dt ds \Delta Q(t) \Delta Q(s) {\cal  \bar{I}}_{\Delta Q, \Delta Q}^{C \to S}(t,s) \cal \\&&e^{\frac{i}{2 \hbar} \int dt C_{SC}(t) \bar{Q}(t) \Delta X^*(t) -\frac{1}{\hbar}\int \int dt ds  \Delta X^*(t) k_r^{B \to C}    (t-s) \Delta X^*(s)} \nonumber
\nonumber \nonumber,,
\end{eqnarray}
where
\begin{eqnarray}
\label{eq:bidqdq}
&&{\cal  \bar{I}}_{\Delta Q, \Delta Q}^{C \to S}(t,s) =4 C_{SC}(t) C_{SC}(s)  \iint^{t_f} du dw \\
&& \nonumber \iiiint^{t_f} dp dp' dr dr'   R_C(p,t)R_C(p',s)R_C(r,u)R_C(r',w)\\&& k_B^{B \to C}(p-r) k_B^{B \to C}(p'-r')   {\cal I}^{B \to C}(u,w) \nonumber
\end{eqnarray}

\section{Subtleties of the Vernon transform in the Fourier domain}
\label{sec:problems}
In this section we focus on just the cavity interacting with the bath.
Suppose that this interaction is time-independent,
that the bath is initially in thermal equilibrium
with respect to the bath Hamiltonian, and that the
Feynman-Vernon kernels 
$(k_I^{b\to c},k_R^{b\to c})$ only depends on the second
time argument. These are the assumptions that 
lead to the Vernon transform in the Fourier domain
\eqref{eq:Vernon-transform-Fourier-side}
\\\\
But are these assumptions consistent?
If the cavity starts in an arbitrary state, the cavity and the bath
will not be in joint equilibrium at the start of the process,
unless the bath-cavity interaction is so weak that they are practically
independent. Classically, the bath would hence tend to equilibrate
conditionally to the initial cavity. The cavity hence
does work on the bath. To this (classical) force must correspond 
a reaction force of the bath on the cavity which is active at the 
beginning of the process. 
Transposed to the quantum domain this means that
the Feynman-Vernon kernels 
$(k_I^{b\to c},k_R^{b\to c})$ 
should also depend on the first time 
argument, at least for some time in the beginning of the 
process. It is known since about two decades that this is the case~\cite{daCosta2000,Ford1985,Ingold2009}.
If the bath is Caldeira-Leggett
and in the limit of very high bath temperature, it is well known
that the evolution of the cavity is Markov, and equivalent to
a (classical) under-damped Langevin equation~\cite{CALDEIRA1983}.
This however does not hold 
at the initial time, where appears a delta-in-time force.
This is what remains in this limit of the general two-time dependence
of $(k_I^{b\to c},k_R^{b\to c})$.
The corresponding contributions to heat and work were discussed by one
of us in \cite{Aurell2017}.
\\\\
Hence, the Vernon transform has to be interpreted with care
when the simplifying assumption is made that the 
Feynman-Vernon kernels 
$(k_I^{b\to c},k_R^{b\to c})$ only depends on the second
time argument.
One approach is to reintroduce a time-dependent system-cavity coupling
which vanishes in the beginning of the process, and which leads 
to Feynman-Vernon kernels that depend on both times.
The output of the Vernon transform, 
\textit{i.e.} the Feynman-Vernon kernels $(k_I^{c\to s},k_R^{c\to s})$,
will then also depend on both times. 
If now the process goes on for a long time and the bath-cavity coupling stays constant
most of that time, the kernels 
$(k_I^{b\to c},k_R^{b\to c})$ will approximately only depend on the
second time argument when the bath-cavity is constant.
This property is inherited by the output kernels.
Focusing for simplicity on just the imaginary kernel
we can thus compare two different Fourier transforms:
\begin{eqnarray}
\hat{k}_I(t,\nu)&=&\int k_I(t,s-t) e^{i\nu s} ds 
\qquad\hbox{"true Fourier"} \nonumber \\
&=& \qquad \int_0^{\infty} e^{i\nu t} k_I(t,\tau)
e^{i\nu \tau}\, d \tau 
\end{eqnarray}
and 
\begin{eqnarray}
\hat{k}'_I(\nu)&=&\int k'_I(s-t) e^{i\nu s} ds 
\qquad\hbox{"truncated Fourier"} \nonumber \\
&=& \qquad \int_0^{\infty} e^{i\nu t} k'_I(\tau)
e^{i\nu \tau}\, d \tau 
\end{eqnarray}
where we assume that $k_I(t,s-t)$ only depends on the first argument for
small and large $t$, and $k'_I(s-t)$ is that $t$-independent function (independent of its
first argument) extended to all $t$.
In both equations we have in the last equality used that 
$k_I(t,s-t)=0$ if $s\leq t$. If the dependence of $k_I(t,\tau)$ on its second argument is essentially
finite range we have $\hat{k}_I(t,\nu) \approx \hat{k}'_I(\nu)$
except for small and large $t$.

\end{document}